\documentclass[nofootinbib,prd]{revtex4}
\usepackage[latin9]{inputenc}
\usepackage{amsmath}
\usepackage{amssymb}
\usepackage{graphicx}

\makeatletter
\@ifundefined{textcolor}{}
{%
 \definecolor{BLACK}{gray}{0}
 \definecolor{WHITE}{gray}{1}
 \definecolor{RED}{rgb}{1,0,0}
 \definecolor{GREEN}{rgb}{0,1,0}
 \definecolor{BLUE}{rgb}{0,0,1}
 \definecolor{CYAN}{cmyk}{1,0,0,0}
 \definecolor{MAGENTA}{cmyk}{0,1,0,0}
 \definecolor{YELLOW}{cmyk}{0,0,1,0}
}

\usepackage{amsfonts}
\providecommand{\U}[1]{\protect\rule{.1in}{.1in}}

\makeatother

\begin{document}

\title{A map between Galilean relativity and special relativity}

\author{Gianluca Mandanici}

\email{Gianluca.Mandanici@unibg.it}

\affiliation{Università degli Studi di Bergamo, Facoltà di Ingegneria, Viale Marconi
5, 24044 Dalmine (Bergamo) Italy.}
\begin{abstract}
A map is discussed that connects, in 1+1 dimensions, Galilei's relativity
to Einstein's special relativity. By means of this map it is possible
to derive special-relativistic formulas from the corresponding Galilean
ones. Beyond being interesting on its own, this map is also significant
with respect to a recent debate on the extension of relativistic symmetries
to the Planck scale (especially in the framework of the so-called
doubly special relativity). The map in fact provides an explicit example
of how can be misleading to interpret a mathematical correspondence
between two relativity schemes as an argument in favor of their physical
equivalence.
\end{abstract}
\maketitle

\section{Introduction}

Special relativity has been proposed in \citep{Einstein:1905ve,Einstein:1905:TKS}
by Einstein with the aim to extend the relativity principle from the
mechanical to the electromagnetic phenomena. The hypothesis of the
invariance of the speed of light $c$ has played a central role in
Einstein's original derivation. However, over the last century, special
relativity has been re-derived starting from many different viewpoints
(see e.g. \citep{1976AmJPh..44..271L,1969JMP....10.1518B,1984AmJPh..52..119M,1994AmJPh..62..157S,2003EJPh...24..315P,2005EJPh...26...33S}).
From the special-relativistic formulas the corresponding Galilean
ones are recovered in the limit in which all the speeds involved are
small when compared to the speed of light, i.e. in the $c^{-1}\rightarrow0$
limit. Whereas, from a mathematical point of view, special relativity
can be seen as a deformation of Galilei's relativity, being $c^{-1}$
the deformation parameter, from a physical point of view the observables
(and their relations) in the two frameworks remain well distinct.
The fact that we are dealing with different theories and with different
physical predictions, unless the $c^{-1}\rightarrow0$ limit is not
taken, is widely accepted. More recently extensions of special relativity
have been considered in literature \citep{AmelinoCamelia:2000mn,AmelinoCamelia:2000ge,Magueijo:2001cr,AmelinoCamelia:2003ex},
motivated by the aim to extend special relativity including a second
invariant scale, $\lambda^{-1}$, eventually connected with the Planck
length/energy. This scale $\lambda^{-1}$ is expected to play a role
in the new transformation formulas between inertial observers analogous
to the role played by $c^{-1}$ in the transition from Galilean relativity
to special relativity. The resulting relativity scheme, also known
as doubly-special relativity, should represent an extension of special
relativity in the same sense in which special relativity has represented
an extension of Galilei's relativity. However reasons of concern have
been raised from certain authors \citep{Rembielinski:2002ic,Jafari:2006rr,Burton:2009ym}
ultimately motivated by the fact that a mathematical map can be found
\citep{Judes:2002bw} that allows to describe doubly-special-relativistic
formulas in terms of the corresponding special-relativistic ones.
The existence of the mentioned map \citep{Judes:2002bw} would prove,
according to these authors (see especially \citep{Jafari:2006rr}),
that we are dealing with the same fundamental theory but written in
terms of different variables. Various objections have already been
raised against this type of arguments \citep{AmelinoCamelia:2010pd,Carmona:2010ze}.
The aim of this paper is to provide a counterexample by showing that
a map between Galilean relativity and special relativity can be found
as well. The map allows us to rewrite special-relativistic formulas
directly from the corresponding Galilean ones, and \textit{viceversa},
without recurring to the $c^{-1}\rightarrow0$ limit. The existence
of such a map represents in our opinion a strong argument against
the deduction of the physical equivalence of two relativity frameworks
motivated by a mathematical correspondence between them. The paper
is organized as follows. In Section II we construct the Galilei-Einstein
map in the coordinate space, in Section III we discuss the map in
the energy-momentum sector, in Section IV we discuss the possibility
to extend the map to 3+1 dimensions, finally, in Section V, we present
our final remarks.

\section{The map in the coordinate space}

We start our derivation of the Galilei-Einstein map remembering that
the infinitesimal actions of the boosts on the coordinate space can
be written, in 1+1 space-time dimensions, in the form
\begin{align}
\Delta x^{\prime} & \simeq\Delta x+\delta_{N}\Delta x,\label{negtrasf}\\
\Delta t^{\prime} & \simeq\Delta t+\delta_{N}\Delta t,
\end{align}
where $\delta_{N}^{\xi}\Delta x=\xi\left\{ N,\Delta x\right\} $ and
$\delta_{N}^{\xi}\Delta t=\xi\left\{ N,\Delta t\right\} $ are the
infinitesimal actions, $\xi$ is the boost parameter and $N$ is the
boost generator. The actions of the boosts on the space-time coordinates
can be expressed in terms of the Poisson brackets 
\begin{align}
\left\{ N,\Delta x_{E}\right\}  & =\Delta t_{E},\label{Lx}\\
\left\{ N,\Delta t_{E}\right\}  & =\Delta x_{E}/c^{2},\label{Lx2}
\end{align}
for the special-relativistic case, and in terms of the brackets 
\begin{align}
\left\{ N,\Delta x_{G}\right\}  & =\Delta t_{G},\label{Gx}\\
\left\{ N,\Delta t_{G}\right\}  & =0,\label{Gx2}
\end{align}
for the Galilean transformations. Here we look for a map between the
Lorentz-Einstein algebra (\ref{Lx})-(\ref{Lx2}) and the Galilei
algebra (\ref{Gx})-(\ref{Gx2}) of the form $\Delta x_{E}=\Delta x_{E}(\Delta x_{G},\Delta t_{G}),$
$\Delta t_{E}=\Delta x_{E}(\Delta x_{G},\Delta t_{G})$. Substituting
in Eqs.(\ref{Lx})-(\ref{Lx2}), and using Eqs.(\ref{Gx})-(\ref{Gx2}),
we find the following system of differential equations: 
\begin{align}
\Delta t_{G}\frac{\partial\Delta x_{E}}{\partial x_{G}} & =\Delta t_{E},\\
\Delta t_{G}\frac{\partial\Delta t_{E}}{\partial\Delta x_{G}} & =\Delta x_{E}/c^{2},
\end{align}
whose solutions can be written as 
\begin{align}
\Delta x_{E} & =F_{1}\left(c\Delta t_{G}\right)\sinh\left(\frac{\Delta x_{G}}{c\Delta t_{G}}\right)+F_{2}\left(c\Delta t_{G}\right)\cosh\left(\frac{\Delta x_{G}}{c\Delta t_{G}}\right),\label{xe-1}\\
\Delta t_{E} & =F_{1}\left(c\Delta t_{G}\right)\cosh\left(\frac{\Delta x_{G}}{c\Delta t_{G}}\right)+F_{2}\left(c\Delta t_{G}\right)\sinh\left(\frac{\Delta x_{G}}{c\Delta t_{G}}\right),\label{te-1}
\end{align}
where $F_{1}$ and $F_{2}$ depend on the Galilei time $c\Delta t_{G}$
alone. For dimensional reasons we can assume that
\begin{eqnarray}
F_{1}\left(c\Delta t_{G}\right) & = & \alpha c\Delta t_{G},\label{eq:F1}\\
F_{2}\left(c\Delta t_{G}\right) & = & \beta c\Delta t_{G},\label{eq:F2}
\end{eqnarray}
with $\alpha$ and $\beta$ dimensionless constants. Substituting
Eqs.(\ref{eq:F1})-(\ref{eq:F2}) into Eqs.(\ref{xe-1})-(\ref{te-1}),
and defining $v_{G}=\Delta x_{G}/\Delta t_{G},$ we get the final
form of the Galilei-Einstein map in the space-time sector
\begin{align}
\Delta x_{E} & =c\Delta t_{G}\left[\alpha\sinh\left(\frac{v_{G}}{c}\right)+\beta\cosh\left(\frac{v_{G}}{c}\right)\right],\label{xe-1-1}\\
\Delta t_{E} & =\Delta t_{G}\left[\alpha\cosh\left(\frac{v_{G}}{c}\right)+\beta\sinh\left(\frac{v_{G}}{c}\right)\right].\label{te-1-1}
\end{align}

The freedom in the choice of the parameters $\alpha$ and $\beta$
can be better understood by looking at the way in which the map acts
on the velocity space. The speed-transformation rule can be obtained
by taking the ratio between both sides of (\ref{xe-1-1}) and (\ref{te-1-1})
getting
\begin{equation}
v_{E}=c\frac{\alpha\sinh\left(v_{G}/c\right)+\beta\cosh\left(v_{G}/c\right)}{\alpha\cosh\left(v_{G}/c\right)+\beta\sinh\left(v_{G}/c\right)}.\label{eq:ve}
\end{equation}

We notice that if we set $\alpha=\beta$ we get $v_{E}=c$ for every
value of $v_{G}$. In this case the whole velocity space of the Galilean
relativity is mapped into the light-cone of the special relativity.
If instead we choose $\beta=0$ we get
\begin{equation}
v_{E}=c\tanh\left(v_{G}/c\right),\label{eq:velimit}
\end{equation}
that maps the Galilean speeds into subluminal Einstein speeds, $v_{E}<c$.
Finally, if we choose $\alpha=0$, we obtain
\begin{equation}
v_{E}=c\coth\left(v_{G}/c\right),
\end{equation}
that corresponds to the case of superluminal propagation in Einstein
relativity, $v_{E}>c$. 

To gain further insight we can consider the action of the map on the
Galilei and the Lorentz-Einstein space-time invariant. Using Eqs.(\ref{xe-1-1})-(\ref{te-1-1})
is easily found that the Lorentz-Einstein (space-time) invariant is
mapped into the Galilei (time) invariant as follows 
\begin{equation}
\Delta x_{E}^{2}-c\Delta t_{E}^{2}=c^{2}\Delta t_{G}^{2}\left(\alpha^{2}-\beta^{2}\right).
\end{equation}

Thus we see that in general if $\left|\alpha\right|>\left|\beta\right|$
two Galilei events whose space and time distances are respectively
$\Delta x_{G}$ and $\Delta t_{G}$ are mapped into space-like distant
Lorentz-Einstein events. If $\left|\alpha\right|=\left|\beta\right|$
the couple of Galilean events is mapped into events that belong to
the same light cone. Finally if $\left|\alpha\right|>\left|\beta\right|$
every couple of Galilean event is mapped into time-like distant space-time
events (see FIG. 1). 
\begin{figure}
\includegraphics[scale=0.4]{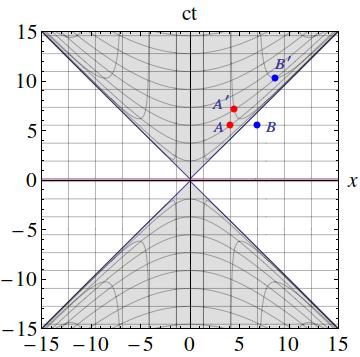}(a)\includegraphics[scale=0.4]{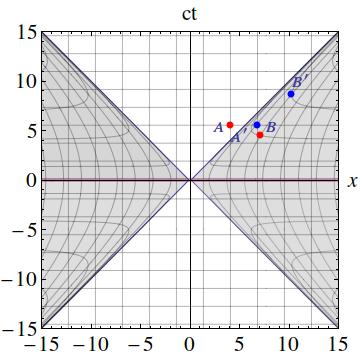}(b)\includegraphics[scale=0.4]{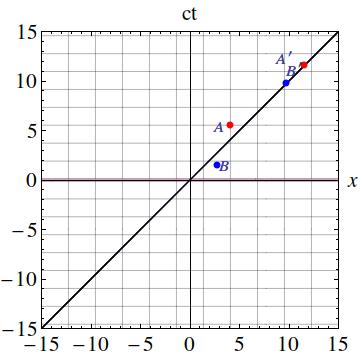}(c)

\label{Figura1}

\caption{(a) $\left|\alpha\right|>\left|\beta\right|,$ Galilei events $A$
and $B$ are mapped into space-like distant events $A'$ and $B'$.
(b) $\left|\alpha\right|<\left|\beta\right|,$ Galilei events $A$
and $B$ are mapped into time-like distant events $A'$ and $B'$.
(c) $\left|\alpha\right|=\left|\beta\right|,$ Galilei events $A$
and $B$ are mapped into light-like distant events $A'$ and $B'$.}
\end{figure}
Equations (\ref{xe})-(\ref{te}) can be easily inverted providing
the inverse map in space-time coordinates:
\begin{eqnarray}
\Delta t_{G} & = & \sqrt{\frac{\Delta x_{E}^{2}/c^{2}-\Delta t_{E}^{2}}{\left(\alpha^{2}-\beta^{2}\right)}},\label{eq:tg}\\
\Delta x_{G} & = & \frac{c}{2}\sqrt{\frac{\Delta x_{E}^{2}/c^{2}-\Delta t_{E}^{2}}{\left(\alpha^{2}-\beta^{2}\right)}}\ln\left[\frac{\left(\alpha-\beta\right)}{\left(\alpha+\beta\right)}\cdot\frac{\Delta t_{E}+\Delta x_{E}/c}{\Delta t_{E}-\Delta x_{E}/c}\right].\label{eq:xg}
\end{eqnarray}

Again taking the ratio between both sides of Eq.(\ref{eq:xg}) and
Eq.(\ref{eq:tg}) one finds
\begin{equation}
v_{G}=\frac{c}{2}\ln\left[\frac{\left(\alpha-\beta\right)}{\left(\alpha+\beta\right)}\cdot\frac{1+v_{E}/c}{1-v_{E}/c}\right],\label{eq:vg}
\end{equation}
that is the inverse of (\ref{eq:ve}). It is straightforward to show
(see Appendix A) using Eqs.(\ref{xe-1})-(\ref{te-1}) that Galilei
transformations are mapped into the Lorentz ones and \textit{viceversa}:
\begin{equation}
\left\{ \begin{array}{l}
x_{G}^{\prime}=x_{G}-v_{G}t_{G}\\
t_{G}^{\prime}=t_{G}
\end{array}\right.\leftrightarrow\left\{ \begin{array}{l}
x_{E}^{\prime}=\gamma(x_{E}-v_{E}t_{E})\\
t_{E}^{\prime}=\gamma(t_{E}-v_{E}/c^{2}x_{E})
\end{array}\right.,
\end{equation}
where, as usual, $\gamma=1/\sqrt{1-v_{E}^{2}/c^{2}}.$ It is also
easy to check the mapping between the formulas of velocity composition:
\begin{equation}
v_{G}^{\prime}=v_{G}+v_{G}^{b}\text{ \ }\leftrightarrow\text{ \ }v_{E}^{\prime}=\frac{v_{E}+v_{E}^{b}}{1+v_{E}v_{E}^{b}/c^{2}},
\end{equation}
that can be derived directly from Eq.(\ref{eq:ve}) and Eq.(\ref{eq:vg}).
Finally we notice that if we request that the map reduces to the identity
in the limit $c^{-1}\rightarrow0$ (i.e. if we ask that $x_{E}=x_{G},$
$t_{E}=t_{G}$) then we find that it must be $\alpha=1$ and $\beta=0,$
so that, in this case, the map (\ref{xe-1-1})-(\ref{te-1-1}) reduces
to the following 
\begin{align}
\Delta x_{E} & =c\Delta t_{G}\sinh\left(\frac{v_{G}}{c}\right),\label{xe}\\
\Delta t_{E} & =\Delta t_{G}\cosh\left(\frac{v_{G}}{c}\right),\label{te}
\end{align}
and the speed to (\ref{eq:velimit}). We also observe that if we define
$\Delta t_{G}=c/a$ and $v_{G}=a\tau$ we get 
\begin{align}
\Delta x_{E} & =\frac{c^{2}}{a}\sinh\left(\frac{a\tau}{c}\right),\label{xe-2}\\
\Delta t_{E} & =\frac{c}{a}\cosh\left(\frac{a\tau}{c}\right),\label{te-2}
\end{align}
that are the coordinates of a Rindler observer.

\section{The map in the energy-momentum space}

Once we have established the procedure in the coordinate space it
is straightforward to extend it to the energy-momentum space. As we
did in the previous section we write the action of the boosts using
the Poisson-bracket notation $\delta_{N}E_{E}=\xi\left\{ N,E_{E}\right\} $
and $\delta_{N}P_{E}=\xi\left\{ N,P_{E}\right\} $, being the infinitesimal
actions given by 
\begin{align}
\left\{ N,P_{E}\right\}  & =E_{E}/c^{2},\label{LA}\\
\left\{ N,E_{E}\right\}  & =P_{E},\label{LA2}
\end{align}
in the special-relativity case, and by 
\begin{align}
\left\{ N,P_{G}\right\}  & =m,\label{GA}\\
\left\{ N,E_{G}\right\}  & =P_{G}.\label{GA2}
\end{align}
in the case of Galilean relativity. Again we look for a map between
the two algebras of the general form $E_{E}=E_{E}(E_{G},P_{G}),$
$P_{E}=P_{E}(E_{G},P_{G})$. From Eqs.(\ref{LA})-(\ref{LA2}) it
follows that the map satisfies the system of equations 
\begin{align}
P_{G}\frac{\partial E_{E}}{\partial E_{G}}+m\frac{\partial E_{E}}{\partial P_{G}} & =P_{E},\\
P_{G}\frac{\partial P_{E}}{\partial E_{G}}+m\frac{\partial P_{E}}{\partial P_{G}} & =E_{E}/c^{2}.
\end{align}

The solutions of the above system are
\begin{align}
P_{E} & =mc\left[\tilde{\alpha}\sinh\left(\frac{P_{G}}{mc}\right)+\tilde{\beta}\cosh\left(\frac{P_{G}}{mc}\right)\right],\label{PE}\\
E_{E} & =mc^{2}\left[\tilde{\alpha}\cosh\left(\frac{P_{G}}{mc}\right)+\tilde{\beta}\sinh\left(\frac{P_{G}}{mc}\right)\right],\label{EE}
\end{align}
where $\tilde{\alpha}$ and $\tilde{\beta}$ are dimensionless constants.
From Eqs.(\ref{PE})-(\ref{EE}) the invariant mass Casimir can be
written as
\[
E_{E}^{2}-c^{2}P_{E}^{2}=m^{2}c^{4}\left(\tilde{\alpha}^{2}-\tilde{\beta}^{2}\right).
\]

If $|\tilde{\alpha}|>|\tilde{\beta}|$ we get the energy-momentum
dispersion relation of a particle (\textit{bradion}) propagating slower
than light ($v_{E}=dE_{E}/dp_{E}<c$). If $|\tilde{\alpha}|<|\tilde{\beta}|$
we get the energy-momentum dispersion relation of a particle (\textit{tachyon})
propagating faster than light ($v_{E}>c$). Finally if $|\tilde{\alpha}|=|\tilde{\beta}|$
we get the energy-momentum dispersion relation of a massless particle
propagating exactly at the speed of light ($v_{E}=c$). As in the
coordinate space, if we request that the map reduces to the identity
(i.e. $E_{E}=E{}_{G},$ $p_{E}=p_{G}$) in the limit $c^{-1}\rightarrow0$,
then we find that it must be $\tilde{\alpha}=1$ and $\tilde{\beta}=0.$
Under this assumption the map reduces to the simpler 
\begin{align}
P_{E} & =mc\sinh\left(\frac{P_{G}}{mc}\right)=mc^{2}\sinh\left(\sqrt{\frac{2E_{G}}{mc}}\right),\label{PE-1}\\
E_{E} & =mc^{2}\cosh\left(\frac{P_{G}}{mc}\right)=mc^{2}\cosh\left(\sqrt{\frac{2E_{G}}{mc}}\right),\label{EE-1}
\end{align}
the inverse map being
\begin{align}
P_{G} & =mc\ln\left[\frac{P_{E}}{mc}+\sqrt{\left(\frac{P_{E}}{mc}\right)^{2}+1}\right],\label{PG}\\
E_{G} & =\frac{mc^{2}}{2}\ln^{2}\left[\frac{E_{E}}{mc^{2}}+\sqrt{\left(\frac{E_{E}}{mc^{2}}\right)^{2}-1}\right].\label{EG}
\end{align}

We have already derived, in space-time coordinates, the transformation
that maps the particle velocity. Here we notice that using the standard
definition of particle speed, $v=dE/dp,$ we can derive the same formula
in the momentum space. In fact we have that
\begin{equation}
v_{E}=\frac{dE_{E}}{dP_{E}}=\frac{dE_{E}}{dE_{G}}\frac{dE_{G}}{dP_{G}}\frac{dP_{G}}{dP_{E}}=v_{G}\dfrac{dE_{E}}{dE_{G}}\left(\dfrac{dP_{E}}{dP_{G}}\right)^{-1},\label{VEgeneral}
\end{equation}
and applying the map of Eqs.(\ref{PE})-(\ref{EE}) to the expression
(\ref{VEgeneral}) we get the desired formula of Eq.(\ref{eq:ve})
in the energy-momentum space. The map of Eqs.(\ref{PG})-(\ref{EG})
connects Galilei algebra to Lorentz-Poincaré algebra. This allows
us to rewrite physical quantities, expressed in the energy-momentum
space and transforming according to Galilei relativity, in terms of
quantities transforming according to special relativity. It is easy
to show that Galilean transformation rules between inertial observers
are mapped into the corresponding special-relativistic ones:
\begin{equation}
\left\{ \begin{array}{l}
p_{G}^{\prime}=p_{G}-v_{G}m\\
E_{G}^{\prime}=E_{G}+v_{G}p
\end{array}\right.\text{ \ }\leftrightarrow\text{ \ }\left\{ \begin{array}{l}
p_{E}^{\prime}=\gamma(p_{E}-v_{E}E_{E})\\
E_{E}^{\prime}=\gamma(E_{E}-v_{E}/c^{2}p_{E}).
\end{array}\right.
\end{equation}

Also the energy-momentum dispersion relations, connected to the Casimir
invariant of the corresponding algebras, are mapped into each other
\begin{equation}
E_{G}=\frac{p_{G}^{2}}{2m}\leftrightarrow E_{E}^{2}=c^{2}p_{E}^{2}+m^{2}c^{4}.\label{eq:QV}
\end{equation}

One can also recover the relation between the energy-momentum and
the speed of a particle. In fact using Eq.(\ref{eq:ve}) and Eq.(\ref{eq:vg})
one finds
\begin{align}
p_{G} & =mv_{G}\text{ \ }\leftrightarrow\text{ \ }p_{E}=\frac{mv_{E}}{\sqrt{1-(v_{E}/c)^{2}}},\\
E_{G} & =\frac{1}{2}mv_{G}^{2}\text{ \ }\leftrightarrow\text{ \ }E_{E}=\frac{mc^{2}}{\sqrt{1-(v_{E}/c)^{2}}}.
\end{align}

Now we are ready to analyze another key issue of the correspondence
between Galilei relativity and Einstein special relativity: the energy-momentum
conservation rule. Energy-momentum conservation rule in special relativity
can be written as
\begin{align}
E_{1E}+E_{2E} & =E_{3E},\\
p_{1E}+p_{2E} & =p_{3E}.
\end{align}

Substituting expressions (\ref{PE})-(\ref{EE}) we find that the
corresponding Galilean conservation rules read
\begin{align}
m_{1L}\cosh\left(\frac{P_{1G}}{m_{1G}c}\right)+m_{2L}\cosh\left(\frac{P_{2G}}{m_{2G}c}\right) & =m_{3L}\cosh\left(\frac{P_{3G}}{m_{3G}c}\right),\label{Econs}\\
m_{1L}\sinh\left(\frac{P_{1G}}{m_{1G}c}\right)+m_{2L}\sinh\left(\frac{P_{2G}}{m_{2G}c}\right) & =m_{3L}\sinh\left(\frac{P_{3G}}{m_{3G}c}\right).\label{Pcons}
\end{align}

Formulas (\ref{Econs})-(\ref{Pcons}) represent fully covariant expressions
in 1+1-dimensional Galilei relativity, for every value of the parameter
$c$, as can be easily verified observing that 
\begin{equation}
\left\{ N,m_{iG}\cosh\left(\frac{P_{iG}}{m_{iG}c}\right)\right\} =\frac{m_{iG}}{c}\sinh\left(\frac{P_{iG}}{m_{iG}c}\right),\qquad\left\{ N,m_{iG}\sinh\left(\frac{P_{iG}}{m_{iG}c}\right)\right\} =\frac{m_{iG}}{c}\cosh\left(\frac{P_{iG}}{m_{iG}c}\right).
\end{equation}

However Eqs.(\ref{Econs})-(\ref{Pcons}) are not the right energy-momentum
conservation laws. The real ones can be obtained eliminating the invariant
parameter, i.e. taking the $c^{-1}\rightarrow0$ limit. In this way
one finds
\begin{align}
m_{1}+m_{2} & =m_{3},\label{eq:conservation_laws_Galilei}\\
p_{1G}+p_{2G} & =p_{3G},\label{eq:Conservation_Law_GAlilei2}
\end{align}
that are the right conservation laws of Galilei relativity. Instead,
if we had started from (\ref{eq:conservation_laws_Galilei})-(\ref{eq:Conservation_Law_GAlilei2})
we would have obtained for the energy-momentum conservation laws of
special relativity
\begin{align}
m_{1}+m_{2} & =m_{3},\label{eq:conservation_laws_Galilei-1}\\
m_{1}c\ln\left[\frac{cP_{1E}+E_{1E}}{m_{1}c}\right]+m_{2}c\ln\left[\frac{cP_{2E}+E_{2E}}{m_{2}c}\right] & =m_{3}c\ln\left[\frac{cP_{3E}+E_{3E}}{m_{3}c}\right],\label{eq:Conservation_Law_GAlilei2-1}
\end{align}
whose covariance can be easily proved noticing that
\begin{equation}
\left\{ N,m_{i}\right\} =0,\qquad\left\{ N,\ln\left(\frac{cP_{iE}+E_{iE}}{m_{i}c}\right)\right\} =\frac{E_{iE}/c+p_{iE}}{cP_{iE}+E_{iE}}=c^{-1}.
\end{equation}

Again one does not obtain the real special-relativistic energy-momentum
conservation laws starting from the real Galilean ones, but again
the energy-momentum conservation laws are compatible with the covariance.
To obtain the real energy-momentum conservation rules of special relativity
one has to start from the deformed, but still covariant, Galilean
mass-momentum conservation rules of Eqs.(\ref{Econs})-(\ref{Pcons}).

\section{On the possibility to extend the map to 3+1 space-time dimensions}

Having constructed the map in 1+1 dimensions one could wonder if a
similar map can be found in 3+1 space-time dimensions as well. We
focus again in the energy-momentum sector but the same procedure can
be easily reproduced in the space-time sector, leading to the same
result. We recall that, in the energy-momentum sector, the relevant
Lorentz-Poincaré algebra in 3+1 dimension can be written as
\begin{alignat}{2}
\left\{ N_{Ei},P_{Ej}\right\}  & =E_{E}c^{-2}\delta_{ij}\qquad\qquad\qquad & \left\{ N_{Ei},E_{E}\right\}  & =P_{Ei}\label{eq:LA1}\\
\left\{ N_{Ei},N_{Ej}\right\}  & =-c^{-1}\epsilon_{ijk}M_{Ek}\qquad\qquad\qquad & \left\{ M_{Ei},M_{Ej}\right\}  & =\epsilon_{ijk}M_{Ek}\label{eq:LA2}\\
\left\{ M_{Ei},P_{Gj}\right\}  & =\epsilon_{ijk}P_{Ek}\qquad\qquad\qquad & \left\{ M_{Ei},N_{Ej}\right\}  & =\epsilon_{ijk}N_{Ek}\label{eq:LA3}\\
\left\{ N_{Ei},E_{E}\right\}  & =P_{Ei} & \left\{ M_{Ei},E_{E}\right\}  & =0,\label{eq:LA4}
\end{alignat}
whereas the Galilei algebra reads
\begin{alignat}{2}
\left\{ N_{Ei},P_{Ej}\right\}  & =m\delta_{ij}\qquad\qquad\qquad & \left\{ N_{Ei},E_{E}\right\}  & =P_{Ei}\label{eq:LA1-1}\\
\left\{ N_{Ei},N_{Ej}\right\}  & =0\qquad\qquad\qquad & \left\{ M_{Ei},M_{Ej}\right\}  & =\epsilon_{ijk}M_{Ek}\label{eq:LA2-1}\\
\left\{ M_{Ei},P_{Gj}\right\}  & =\epsilon_{ijk}P_{Ek}\qquad\qquad\qquad & \left\{ M_{Ei},N_{Ej}\right\}  & =\epsilon_{ijk}N_{Ek}\label{eq:LA3-1}\\
\left\{ N_{Ei},E_{E}\right\}  & =P_{Ei} & \left\{ M_{Ei},E_{E}\right\}  & =0.\label{eq:LA4-1}
\end{alignat}

As in the 1+1 dimensional case we look for transformations of the
type $E_{E}=E_{E}(E_{G},\vec{P}_{G}),$ $\vec{P}_{E}=\vec{P}_{E}(E_{G},\vec{P}_{G})$,
that substituted in (\ref{eq:LA1}) with the help of (\ref{eq:LA1-1})
furnish the following system of differential equations 
\begin{align}
\frac{\partial P_{Ej}}{\partial P_{Gi}}m+\frac{\partial P_{Ej}}{\partial E_{G}}P_{Gi} & =\delta_{ij}E_{E}c^{-2},\label{F1}\\
\frac{\partial E_{E}}{\partial P_{Gi}}m+\frac{\partial E_{E}}{\partial E_{G}}P_{Gi} & =P_{Ei}.\label{F2}
\end{align}

We notice here that due to the rotational symmetry one can write the
Lorentz-Einstein momentum as a function of the Galilei energy and
momentum in the form $\vec{P}_{E}=\vec{P}_{G}F(E_{G},P_{G}^{2})$,
where $F$ is a rotationally-invariant function. Substituting $\vec{P}_{E}$
in (\ref{F1}) we get
\begin{equation}
\delta_{ij}(Fm-E_{E}c^{-2})+P_{Gj}P_{Gi}\left(2\frac{\partial F}{\partial P_{G}^{2}}m+\frac{\partial F}{\partial E_{G}}\right)=0.\label{eq:3DR_base}
\end{equation}

Considering the rotational properties of this last equation, and that
it must hold also for $i\neq j$, it follows that it must be
\begin{equation}
\left(2\frac{\partial F}{\partial P_{G}^{2}}m+\frac{\partial F}{\partial E_{G}}\right)=0,\label{nullo}
\end{equation}
that means that $F$ must have the form
\begin{equation}
F(E_{G},P_{G}^{2})=F\left(E_{G}-\frac{P_{G}^{2}}{2m}\right).
\end{equation}

From (\ref{eq:3DR_base}) and (\ref{nullo}) also follows that
\begin{equation}
E_{E}=mc^{2}F\left(E_{G}-\frac{P_{G}^{2}}{2m}\right).
\end{equation}

Substituting $E_{E}$ in Eq.(\ref{F2}) one finds $P_{E}=0$ which
implies that also $F(E_{G},P_{G}^{2})=0$ \ and $E_{G}=0$. Thus
we cannot find a extension of the map (\ref{PE})-(\ref{EE})\ for
the 3+1-dimensional case.

\section{Final remarks and implication for the triviality of doubly special
relativity}

We have argued that Galilei transformations in 1+1 dimensions can
be mapped into Lorentz-Einstein transformations, and \textit{viceversa},
for any (non-zero) value of the deformation parameter $c$. We have
also found that the map depends on two further parameters whose values
tune the mapping of Galilei events into space-like, time-like or light-like
Lorentz-Einstein events. A similar result can be found for the energy-momentum
sector. In this case, depending on the values of the parameters involved,
we can have Galilei particles mapped into bradion-like, photon-like
or tachyons-like Lorentz-Einstein particles. The map also acts on
the velocity space, on the space-time invariant, on the energy-momentum
dispersion relation, on the energy-momentum conservation rules and
on all the other relevant physical quantities. With respect to the
energy-momentum conservation rules we have shown that the Galilei
mass-momentum conservation laws are not immediately mapped into the
special-relativistic energy-momentum conservation laws. Rather they
are mapped into Eqs.(\ref{eq:conservation_laws_Galilei-1})-(\ref{eq:Conservation_Law_GAlilei2-1})
that are still covariant expressions in special relativity but are
not right ones. In order to get the right ones it is necessary to
consider a proper combinations of the Galilei mass-momentum conservation
laws (Eqs.(\ref{Econs})-(\ref{Pcons})). We have derived the map
at a classical level, however it is rather easy, by means of the substitutions
$E\rightarrow i\hbar\partial/\partial t$, $p\rightarrow i\hbar\partial/\partial x$
and Eqs.(\ref{eq:QV}), to find a quantum version of the map changing
a Schrödinger equation into a Klein-Gordon equation, or \textit{viceversa}.
The existence of this 1+1-dimensional map does not mean that 1+1-d
Galilei relativity is equivalent to 1+1-d special relativity. The
observables in the two schemes are different and so are the relations
between them (the physical laws) and, as a consequence, the physical
predictions, at least in the regime in which the map does not reduce
to the identity. The existence of the map only allows us to write
physical laws in one scheme in terms of the physical laws of the image
theory. As we have already said, within the proposals of considering
extensions of special relativity at the Planck scale \citep{AmelinoCamelia:2000mn,AmelinoCamelia:2000ge,Magueijo:2001cr,AmelinoCamelia:2003ex}
also similar maps have been found \citep{Judes:2002bw} connecting
these models to special relativity. The existence of such maps, analogous
to the map we have analyzed here, has led various authors to consider
doubly special relativity as a simple rewriting of special relativity
in term of nonstandard variables. Our result, that special relativity
can be obtained by a nonlinear map from Galilei relativity, shows that
special relativity could be considered as a deformation of Galilei
relativity in the same way. Thus if one accepts, as we do, that Galilei
relativity and special relativity have profound physical differences,
one has to conclude that these maps can account for the richness of
the target model rather than showing its triviality.

\section*{Acknowledgments}

The author thanks Giovanni Amelino-Camelia and Remo Garattini for
the useful discussions.

\section*{APPENDIX A}

Here we assume $\alpha=1$ and $\beta=0$, but the same calculation
can be carried out for $\alpha=0$ and $\beta=1,$ and then easily
generalized to arbitrary values of $\alpha$ and $\beta$. Starting
from the Lorentz boosts
\begin{equation}
\left\{ \begin{array}{l}
\Delta x_{E}^{\prime}=\gamma(\Delta x_{E}-v_{E}\Delta t_{E})\\
\Delta t_{E}^{\prime}=\gamma(\Delta t_{E}-v_{E}/c^{2}\Delta x_{E})
\end{array}\right.
\end{equation}
we substitute the expressions (\ref{xe-1-1})-(\ref{te-1-1}) obtaining
\begin{equation}
\left\{ \begin{array}{c}
c\Delta t_{G}^{\prime}\sinh\left(\dfrac{\Delta x_{G}^{\prime}}{c\Delta t_{G}^{\prime}}\right)=\dfrac{1}{\sqrt{1-\tanh^{2}(v_{G}/c)}}c\Delta t_{G}\left[\sinh\left(\dfrac{\Delta x_{G}}{c\Delta t_{G}}\right)-\tanh(v_{G}/c)\cosh\left(\dfrac{\Delta x_{G}}{c\Delta t_{G}}\right)\right]\\
\Delta t_{G}^{\prime}\cosh\left(\dfrac{\Delta x_{G}^{\prime}}{c\Delta t_{G}^{\prime}}\right)=\dfrac{1}{\sqrt{1-\tanh^{2}(v_{G}/c)}}\Delta t_{G}\left[\cosh\left(\dfrac{\Delta x_{G}}{c\Delta t_{G}}\right)-\tanh(v_{G}/c)\sinh\left(\dfrac{\Delta x_{G}}{c\Delta t_{G}}\right)\right]
\end{array}.\right.\label{eq:Last}
\end{equation}

Taking the square of both sides of (\ref{eq:Last}) and subtracting
from the first equation the second, after some algebraic manipulations
we find
\begin{equation}
\left\{ \begin{array}{l}
\Delta t_{G}^{^{\prime}}=\Delta t_{G}\\
\cosh\left(\dfrac{\Delta x_{G}^{\prime}}{c\Delta t_{G}^{\prime}}\right)=\cosh\left(\dfrac{\Delta x_{G}}{c\Delta t_{G}}-v_{G}/c\right)
\end{array}\right.\rightarrow\left\{ \begin{array}{l}
\Delta t_{G}^{^{\prime}}=\Delta t_{G}\\
\Delta x_{G}^{\prime}=\Delta x_{G}-v_{G}\Delta t_{G},
\end{array}\right.
\end{equation}
that are just the finite Galilei boosts. Repeating the passages in
the reverse order we get Lorentz boosts starting from the Galilei
ones.

\bibliographystyle{ieeetr}

\end{document}